\documentclass[aps,twocolumn,prc]{revtex4-1}

\usepackage{hyperref,amsmath,graphicx,url}
\usepackage[latin1]{inputenc}

\usepackage{xcolor}

\newcommand{\gev}{{\rm \, GeV}}
\newcommand{\mev}{{\rm \, MeV}}

\begin{document}

\title{Deuteron production in AuAu collisions at $\sqrt{s_{NN}} = 7 - 200$ GeV via pion catalysis}

\author{D. Oliinychenko$^1$}
\author{C. Shen$^{2,3}$}
\author{V. Koch$^1$}

\address{1 Cyclotron Rd., Berkeley, California, US, 94720}
\address{2 Institute for Nuclear Theory, University of Washington, Box 351550, Seattle, WA, 98195, USA}
\address{3 Department of Physics and Astronomy, Wayne State University, Detroit, Michigan 48201, USA}
\address{4 RIKEN BNL Research Center, Brookhaven National Laboratory, Upton, NY 11973, USA}

\begin{abstract}
We study deuteron production using no-coalescence hydrodynamic + transport simulations of central AuAu collisions at $\sqrt{s_{NN}} = 7 - 200 \gev$. Deuterons are sampled thermally at the transition from hydrodynamics to transport, and interact in transport dominantly via $\pi p n \leftrightarrow \pi d$ reactions. The measured proton, Lambda, and deuteron transverse momentum spectra and yields are reproduced well for all collision energies considered. We further  provide a possible explanation for  the measured minimum in the energy dependence of the  coalescence parameter, $B_2(\sqrt{s_{NN}})$ as well as for the difference between $B_2(d)$ for deuterons and that for anti-deuterons, $B_2(\bar{d})$.
\end{abstract}

\maketitle

\section{Introduction}

Heavy ion collisions are often called ``Little Bang'' due to a rapid expansion, cooling, and a sequence of freeze-outs reminiscent of the evolution of the early Universe. Another common feature of the Little and Big Bangs is nucleosynthesis, or production of light nuclei. The Big Bang nucleosynthesis for deuterons primarily occurred via $pn \leftrightarrow d\gamma$ reaction. In relativistic heavy ion collisions this reaction does not have sufficient time to create the observed amount of deuterons, which follows from its small cross section and the typical time of collision $\sim  10^{-23} - 10^{-22}$~s. Here other reactions are at work, depending on collision energy and rapidity region. In particular, we have previously suggested that the pion catalysis reaction $\pi p n \leftrightarrow \pi d$ plays the dominant role in deuteron production at $\sqrt{s_{NN}} = 2760 $ GeV in the mid-rapidity region \cite{Oliinychenko:2018ugs,Oliinychenko:2018odl}. In the present paper we shall argue that the same reaction is still the most important  one down to collision energies of $\sqrt{s} = 7\gev$ for deuteron production at mid-rapidity.

We note, however, that the two most popular models of deuteron production --- thermal \cite{Siemens:1979dz,Andronic:2010qu,Andronic:2012dm,Cleymans:2011pe,Oliinychenko:2016dtb} and coalescence \cite{Kapusta:1980zz,Sato:1981ez,Gutbrod:1988gt,Mrowczynski:1992gc,Csernai:1986qf,Polleri:1997bp,Mrowczynski:2016xqm,Bazak:2018hgl,Sun:2016rev,Dong:2018cye,Sun:2018jhg,Scheibl:1998tk} models -- do not need to explicitly involve any particular reactions, although implicitly interactions are assumed which either lead to equilibration (thermal model) or are responsible for the deuteron formation (coalescence). The thermal model postulates that light nuclei are created from a fireball in chemical equilibrium with hadrons. At the chemical freeze-out the reactions that change hadron yields cease and hadrons only continue to collide (quasi-)elastically. These collisions change the momentum distributions, but do not change the yields. Thus, for hadrons the chemical freeze-out temperature, $T_{CFO}$, which is determined from hadron yields, is larger than the kinetic temperature $T_{KFO}$ which is extracted from the momentum spectra, $T_{CFO} > T_{KFO}$. This picture is supported by the fact that the yield-changing reactions typically have smaller cross sections, so they cease earlier during the expansion of the fireball. Deuteron yields and spectra are consistent with the same $T_{CFO}$ and $T_{KFO}$ for nuclei as for hadrons \cite{Adam:2015vda}. This means that they have to be colliding with other particles between chemical and kinetic freeze-out. However the 2.2 MeV binding energy of deuterons is much smaller than $T_{CFO} \approx 150$ MeV or $T_{KFO} \approx 110$ MeV. Simple intuition tells that a deuteron must be easily destroyed at such temperatures. Due to this intuition light nuclei in heavy ion collisions were called ``snowballs in hell'' \cite{cern_courier}, where light nuclei would be ``snowballs'' and the fireball of the heavy ion collisions is referred to as ``hell''. However, this simple intuition fails in two ways. Firstly, even despite the small binding energy of a deuteron, the elastic cross section of $d + \pi \to d + \pi$ reaches as high as 70 mb at the kinetic energies of pion and proton corresponding to temperatures of 100--150 MeV (see Fig. 1 of \cite{Oliinychenko:2018ugs}). One assumes that a thermal pion at $T\approx 150 \mev$ should easily break up a deuteron, but in $\sigma_{\pi d}^{elastic}/\sigma_{\pi d}^{total} \approx 1/4$ of all $\pi + d$ collisions this does not happen: instead, the pion excites one of the nucleons, which subsequently de-excites emitting a pion back while leaving the deuteron  intact. Secondly, the  inelastic reactions that destroy deuterons ($d + X \leftrightarrow p + n + X$, where $X$ is an arbitrary hadron) have backreactions that can also create deuterons. We have shown in \cite{Oliinychenko:2018ugs,Oliinychenko:2018odl} that for Pb+Pb collisions at  $\sqrt{s_{NN}} = 2.76 $ TeV deuteron creation and destruction occur at approximately equal rates between $T_{CFO}$ and $T_{KFO}$. This mechanism, thus,  resolves the ``snowballs in hell'' paradox.  It justifies calculating the deuteron yield in the hadron resonance gas model at $T_{CFO}$ while determining the  deuteron momentum spectrum in a blast wave model at $T_{KFO}$.

In contrast to the thermal model, coalescence models postulate that deuterons are produced from nucleons at the kinetic freeze-out. Nucleons coalesce into a deuteron in case they are close in the phase space. Coalescence can be understood either as an interaction of two off-shell nucleons forming a deuteron, or as a consequence of a reaction like $pn \to d \gamma$, or $pn \to d \pi$, or $p n \to X d$, or $X p n \to X d$, where the concrete species of the particle $X$ is not important for coalescence models. Sometimes not only $pn \to d$ coalescence is considered, but also coalescence of two protons or two neutrons to a deuteron, see for example \cite{Sombun:2018yqh}. This  implies charge exchange reactions like $\pi^{-} p p \leftrightarrow \pi^{0} d$, $\pi^0 p p \leftrightarrow \pi^+ d$, $\pi^0 n n \leftrightarrow \pi^- d$, $\pi^+ n n \leftrightarrow \pi^0 d$. Thinking about coalescence in terms of reactions is not the only possible approach, but it appears to be fruitful, for example $\Lambda p \leftrightarrow d K$ reaction is possible, therefore one may consider coalescence of $\Lambda$ and $p$ to deuteron, which to our knowledge has, so far, not been taken into account. 

Thermal and coalescence models are considered to be opposite and conflicting scenarios for deuteron production. Indeed, the thermal model postulates deuteron production at chemical freeze-out and coalescence postulates production at kinetic freeze-out. In the thermal model nucleons from resonances do not contribute to deuterons, while in coalescence all nucleons, including the ones from resonance decays, are able to produce deuterons. In coalescence the spatial extend of deuteron wavefunction relative to the size of colliding system matters, while in the thermal model it does not. Despite these differences, we are able to accommodate the core ideas of both thermal and coalescence models in our dynamical simulation in the following way: We use relativistic hydrodynamics to simulate the locally equilibrated fireball evolution until the chemical freeze-out. Then we switch to particles and allow them to rescatter using a hadronic cascade. To this end we sample deuterons and all other hadrons according to the local temperature and chemical potential of the switching hypersurface, which in our work is controlled by a certain value of the energy density, $\epsilon_{sw}$. This transition from hydrodynamic to transport is often referred to as particlization. The deuteron yield at this moment is the yield one would obtain in the thermal model, where the volume $V$ is determined by the particlization hypersurface. The majority of these initial (thermal) deuterons is  destroyed in the subsequent kinetic evolution, but at the same time the new ones are created, so that the average yield remains approximately the same. Like in the coalescence model deuterons that finally survive are mostly, although not exclusively, those which are created very late in the hadronic evolution, and thus do not experience any more  collisions. Also our rate of deuteron production has a large peak at low relative momentum between nucleons, which means they have to be close in the phase space to make a deuteron, as the coalescence model postulates.

Conceptually our approach here is the same as in our previous study at 2760 GeV~ \cite{Oliinychenko:2018ugs,Oliinychenko:2018odl}. The key difference is that at lower energies one has to account for the evolution of the net-baryon current, which does not vanish anymore. This requires additional equations for baryon current conservation in the hydrodynamic simulation and the specification of initial condition for the baryon density. These extensions are presently not available in the \texttt{CLVisc} code we used in previous study. Therefore, for the  present study the \texttt{MUSIC} code for the 3D hydrodynamical evolution of the background medium together with a geometric-based initial condition, which had already been tuned to reproduce hadron spectra at the considered energy range \cite{Shen:2020jwv}.
In our previous study we found that due to reactions like $\pi pn \leftrightarrow \pi d$ the deuteron yields and spectra and intimately related to those of the protons. Therefore, our primary concern is a good description of the measured proton spectra. For the considered collision energies ($\sqrt{s_{NN}} \ge 7$ GeV) light nuclei production is merely a small perturbation over the space-time evolution of baryon density. For example, the d/p ratio at 7.7 GeV at midrapidity is around 0.03, and at higher energies this ratio is even smaller. Therefore, one may view any dynamical model of light nuclei production as a combination of a ``background'' of expanding fireball with nucleon density evolving in space-time, and of a ``perturbation'' that acts on this background and creates (and possibly disintegrates) nuclei. No matter how detailed and realistic the nuclei production model is, the overall precision cannot be better than that of the background model. That is why we pay  attention to fitting proton yields and spectra well. It turned out that a precise account of weak decays is important, so we also make sure that we reproduce the yields of $\Lambda$ baryon.

As we shall show, applying our model of deuteron production to Au+Au collisions at energies from 7 to 200 GeV we observe that we are able to describe the measured deuteron spectra and yields using the same reactions $\pi  p n  \leftrightarrow \pi d$, $N pn \leftrightarrow N d$, $NN \leftrightarrow \pi d$ with the same cross sections that we used to describe the yields, spectra and flow at 2760 GeV \cite{Oliinychenko:2018ugs,Oliinychenko:2018odl}. The work is organized as follows: in Section \ref{sec:methodology} we explain the details of our simulation, and in Section \ref{sec:results_deuteron} we present and discuss the resulting proton, $\Lambda$, and deuteron spectra, yields, and reaction rates relevant for deuteron production. Finally, we explore the role of the correction from the weak decays feeddown to protons in the context of various observables involving deuterons.

\section{Hydrodynamics + transport simulation methodology} \label{sec:methodology}

To simulate the full evolution of a system created in heavy ion collisions we employ a hybrid relativistic hydrodynamics + hadronic transport approach. Hydrodynamics is applied at the earlier stage of collision, where the density is high and, therefore, hadrons cannot be treated as individual particles. The transition from the quark-gluon plasma to a hadron gas is handled implicitly via the equation of state used in the hydrodynamics. Hadronic transport is applied at the later stage of collision, when the fireball is dilute enough so that mean free paths of the particles are larger than their Compton wavelengths.

\subsection{Initial state}
The initial conditions and hydrodynamic simulations are based on the work in Ref.~\cite{Shen:2020jwv}.  In our study the main role of this initial state is that it allows to reproduce the measured proton phase-space distributions and midrapidity yields at different collision energies. Here we mention some features of the initial state, and the full description can be found in Ref.~\cite{Shen:2020jwv}. We start simulations by generating event-averaged initial energy density and net baryon density profiles for hydrodynamics at a constant proper time $\tau = \sqrt{t^2 - z^2} = \tau_0$, where $z$ is the coordinate along the collision axis. The proper time $\tau_0$ is a function of collision energy, which is chosen to be slightly longer than the overlapping time that would take nuclei to pass through each other in absence of interactions $\tau_{\rm overlap} = \frac{2R}{\sqrt{\gamma_L^2 - 1}}$, where $\gamma_L = \frac{\sqrt{s_{\mathrm{NN}}}}{2 m_N}$. The values of $\tau_0$ were chosen such that the the hydrodynamics + hadronic transport simulations can reasonably reproduce the measured mean transverse momentum of identified particles \cite{Shen:2020jwv}.
The initial energy-momentum tensor is assumed to have a diagonal ideal-fluid form $T^{\mu\nu} = (\epsilon + p)u^{\mu}u^{\nu} - p g^{\mu\nu}$. The initial baryon current is also assumed to have an ideal-fluid form, $j^{\mu} = n_B u^{\mu}$. 
At $\tau = \tau_0$, Bjorken flow is assumed: $u^\mu = (\cosh \eta_s, 0, 0, \sinh \eta_s)$.
Based on the local energy and momentum conservation, the local rest frame energy-density $\epsilon(x,y,\eta_s)$ profile in case of our symmetric Au+Au collision system is parametrized as described in \cite{Shen:2020jwv}:
\begin{eqnarray}
 && e (x, y, \eta_s; y_\mathrm{CM})  = \mathcal{N}_e(x, y) \nonumber \\
    && \!\!\!\!\!\quad \times \exp\left[- \frac{(\vert \eta_s \vert  - \eta_0)^2}{2\sigma_\eta^2} \theta(\vert \eta_s \vert - \eta_0)\right],
\end{eqnarray}
where the normalization factor $\mathcal{N}_e(x, y)$ is,
\begin{eqnarray}
   &&\mathcal{N}_e (x, y) = \frac{m_N \sqrt{T_A^2 + T_B^2 + 2 T_A T_B \cosh(2y_\mathrm{beam})}}{2 \sinh(\eta_0) + \sqrt{\frac{\pi}{2}} \sigma_\eta e^{\sigma_\eta^2/2} C_\eta}  \\
   &&C_\eta = e^{\eta_0}\mathrm{erfc}\left(-\sqrt{\frac{1}{2}} \sigma_\eta\right)  + e^{-\eta_0}\mathrm{erfc}\left(\sqrt{\frac{1}{2}} \sigma_\eta\right) \\
   &&y_\mathrm{beam} = \mathrm{arccosh}[\sqrt{s_\mathrm{NN}}/(2 m_N)]
\end{eqnarray}
and $T_{A}(x, y)$ and $T_B(x, y)$ are the nuclear thickness functions for the incoming projectile and target nucleus. Although our colliding system is symmetric, the local nuclear thickness functions $T_{A}(x, y) \neq T_{B}(x, y)$ at a nonzero impact parameter. For the initial baryon density profile  $n_B(x,y,\eta_s)$, we use parametrization as was in Refs.~\cite{Shen:2020jwv, Denicol:2018wdp},
\begin{equation}
    n_B(x, y, \eta_s) = \frac{1}{\tau_0}[T_A(x, y) f^{A}_{n_B} (\eta_s) + T_B(x, y) f^{B}_{n_B} (\eta_s)]. 
\end{equation}
Here the longitudinal profiles are parametrized with asymmetric Gaussian profiles,
\begin{eqnarray}
    &&f^{A}_{n_B} (\eta_s) = \mathcal{N}_{n_B} \left\{\theta(\eta_s - \eta_{B,0}) \exp\left[ - \frac{(\eta_s - \eta_{B,0})^2}{2\sigma_{B, \mathrm{out}}^2} \right] \right. \nonumber \\
    && \qquad \qquad + \theta(\eta_{B,0} - \eta_s) \exp\left[ - \frac{(\eta_s - \eta_{B,0})^2}{2\sigma_{B, \mathrm{in}}^2} \right] \bigg\}
    \label{eq:nBprofr}
\end{eqnarray}
and
\begin{eqnarray}
    && f^{B}_{n_B} (\eta_s) = \mathcal{N}_{n_B} \bigg\{\theta(\eta_s + \eta_{B,0}) \exp\left[ - \frac{(\eta_s + \eta_{B,0})^2}{2\sigma_{B, \mathrm{in}}^2} \right] \nonumber \\
    && \qquad \qquad + \theta(- \eta_{B,0} - \eta_s) \exp\left[ - \frac{(\eta_s + \eta_{B,0})^2}{2\sigma_{B, \mathrm{out}}^2} \right] \bigg\}.
    \label{eq:nBprofl}
\end{eqnarray}
The normalization factor $\mathcal{N}_{n_B}$ is chosen such that $\int d\eta_s f^{A/B}_{n_B} (\eta_s) = 1$.
All the model parameters are specified in Table 1 of Ref.~\cite{Shen:2020jwv}.

\subsection{Hydrodynamics} To calculate the hydrodynamic evolution numerically we use the open-source hydrodynamic code, \texttt{MUSIC v3.0} \cite{Schenke:2010nt, Schenke:2011bn, Paquet:2015lta, Denicol:2018wdp, MUSIC}. The hydrodynamics equations include energy-momentum and baryon number conservation, equation of state $p=p(\epsilon, n_B)$, and Israel-Stewart-type relaxation equations for the viscous stress tensor. In this work we include only shear viscous corrections, while bulk viscous corrections and baryon diffusion are neglected. We used a lattice QCD based ``NEOS-BSQ'' equation of state $p=p(\epsilon, n_B)$ described in Ref.~\cite{Monnai:2019hkn}. It smoothly interpolates between the equation of state at high temperature from lattice QCD \cite{Bazavov:2014pvz} and hadron resonance gas at low temperature. Higher-order susceptibilities were used to extend this EoS to finite baryon chemical potential as a Taylor expansion.
This equation of state explicitly imposes strangeness neutrality, $n_S=0$, and constrains the ratio of the local net charge density to net baryon density to that of the colliding nuclei,  $n_Q/n_B = 0.4$. A temperature and chemical potential dependent specific shear viscosity $\eta/s(T, \mu_B)$ was included. The detailed parameterization in given in Fig.4 in Ref.~\cite{Shen:2020jwv}. The hydrodynamic equations are evolved in $\tau$ until all computational cells reach energy density below $\epsilon_{sw}$. A particlization hypersurface of constant ``switching'' energy density $\epsilon_{sw}$ is constructed and its normal four-vectors $d\sigma_{\mu}$ are computed as described in \cite{Huovinen:2012is}. The value of $\epsilon_{sw} = 0.26$ GeV/fm$^3$ is adjusted to fit bulk observables in \cite{Shen:2020jwv}. This value results in a good simultaneous fit of pion, kaon, and net proton observables across the range of energies that we study. However, at 39 GeV and above the proton yields at midrapidity are somewhat underestimated. To fine-tune proton yields at midrapidity we take higher $\epsilon_{sw}$ at highest STAR energies: $\epsilon_{sw} = 0.35$ GeV/fm$^3$ at $\sqrt{s_{NN}} = 39$ GeV, $\epsilon_{sw} = 0.45$ GeV/fm$^3$ at $\sqrt{s_{NN}} = 62.4$ GeV, and $\epsilon_{sw} = 0.5$ GeV/fm$^3$ at $\sqrt{s_{NN}} = 200$ GeV. In principle the fine-tuning of proton yields can be performed by adjusting other parameters as well, but tuning $\epsilon_{sw}$ was the simplest solution, because pion and kaon yields are known to be rather insensitive to $\epsilon_{sw}$, in contrast to proton and Lambda yields (see Fig. 8 in \cite{Monnai:2019hkn}). Fine-tuning of $\epsilon_{sw}$ allowed us to reproduce proton and Lambda yields slightly better, at the cost of reproducing anti-proton yields slightly worse.

Performing particlization on the obtained hypersurfaces and allowing the generated hadrons to subsequently re-scatter via a hadronic transport approach, one obtains a good description of the measured pion, kaon, and proton yields, transverse momentum and rapidity spectra, and flow $v_2$~\cite{Shen:2020jwv}. The particlization is a standard grand-canonical Cooper-Frye particlization, conducted by the \texttt{iSS} sampler v1.0, which was described and tested in \cite{Shen:2014vra} and is available publicly at \cite{url:iSS}.

\subsection{Transport simulation}

The hydrodynamic evolution and particlization are followed by a hadronic afterburner, where particles are allowed to scatter and decay. For this purpose we use the relativistic transport code \texttt{SMASH} \cite{Weil:2016zrk}, version 1.7, in the cascade mode (= no mean field potentials). Transport simulation is initialized from particles at particlization.
Then the whole system is propagated from action (collision or decay) to action, using a list of actions sorted by time in the computational frame. A collision is added to the list by the geometrical criterion, $\pi d_{tr}^2 < \sigma$, where $d_{tr}$ is the transverse distance in the center of mass frame of colliding particles, and $\sigma$ is the total cross section. The collision time is defined as a time of the closest approach, the decay time is randomly drawn from the exponential distribution, which takes time dilation into account.
For new particles produced by actions, we search for their subsequent collisions and decays, and merge the found ones into the sorted list of actions. This timestep-less collision finding in \texttt{SMASH 1.7} is an improvement compared to the timestep-based one in \texttt{SMASH 1.0}, described in the original publication \cite{Weil:2016zrk}, see the comparison and thorough testing in \cite{Ono:2019ndq}. The end time of our transport simulation is set to 100 fm/c, when the system is already too dilute to sustain even the reactions with very large cross sections, such as deuteron formation, as evident from Fig. \ref{fig:reactions}.

Possible reactions in \texttt{SMASH} include: elastic collisions, resonance formation and decay, $2 \to 2$ inelastic reactions such as $NN \to N\Delta$, $NN \to N N^*$, $NN \to N\Delta^*$ ($N^*$ and $\Delta^*$ denote all nucleon- and delta-resonances), and strangeness exchange reactions; string formation and immediate decay into multiple hadrons. The \texttt{SMASH} resonance list comprises most of the hadron resonances listed in the Particle Data Group collection \cite{Patrignani:2016xqp} with pole mass below 2.6 GeV. The main update relevant for this study since the publication \cite{Weil:2016zrk} is the high-energy hadronic interactions via string formation, in particular baryon-antibaryon annihilations. All the reactions, except the ones with strings, obey the detailed balance principle. The implementation of hadronic interactions in \texttt{SMASH} is described in detail in \cite{Weil:2016zrk}, while \cite{Steinberg:2018jvv} is devoted specifically to reactions involving strangeness. Soft string formation and fragmentation are similar to the the \texttt{UrQMD} code \cite{Bass:1998ca} and described in detail in \cite{Mohs:2019iee}. The main difference to the \texttt{UrQMD} implementation is that the Lund fragmentation functions from Pythia 8 \cite{Sjostrand:2007gs} are employed for string fragmentation.

The deuteron treatment is the same as in \cite{Oliinychenko:2018ugs}, with the same reactions and the same cross sections following the detailed balance principle: $\pi d \leftrightarrow \pi np$, $N d \leftrightarrow N np$, $\bar{N} d \leftrightarrow \bar{N} np$, $\pi d \leftrightarrow NN$ reactions, elastic $\pi d$, $N d$ and $\bar{N} d$ and all of their CPT-conjugates. The latter produce and destroy anti-deuterons, and one can observe their role in Fig. \ref{fig:dn_dy}, where anti-deuteron yields are reproduced rather well.
The deuteron is modelled as a pointlike particle, as it is done in  \cite{Oliinychenko:2018ugs,Oh:2009gx,Danielewicz:1991dh,Longacre:2013apa}. Treating deuterons as pointlike particles is only justified, when the mean free path is at least twice larger than the deuteron size. In our simulation this condition is fulfilled only after $t \approx 10-20$ fm/c depending on the collision energy. At earlier time our deuterons are not defined as particles and should be understood as correlated nucleon pairs. The reactions $\pi d \leftrightarrow \pi np$, $N d \leftrightarrow N np$, $\bar{N} d \leftrightarrow \bar{N} np$, are implemented in two steps using a fake $d'$ resonance:
\begin{eqnarray}
p n \leftrightarrow d' \\
d' X \leftrightarrow d X \,,
\end{eqnarray}
where $X$ can be a pion, a  nucleon, or an anti-nucleon. The $d'$ pole mass is taken to be $m_{d'} = m_{d} + 10$ MeV and width is $\Gamma_{d'} = 100$ MeV, the spin of $d'$ is assumed to be 1. The motivation for this width is to have the $d'$ lifetime close to the time that proton and neutron spend flying past each other. The deuteron disintegration cross sections reach 200 mb, see Figs. 1-2 of \cite{Oliinychenko:2018ugs}. As a consequence of the detailed balance relations and sharp $d'$ spectral function, the $\pi d'$ and $N d'$ cross sections are even larger, reaching up to 1500 mb peak values. To find these reactions correctly, the default collision finding cutoff in \texttt{SMASH} is increased to 2000 mb --- the same value that was used in \cite{Oliinychenko:2018ugs}. This cutoff is the only change to the \texttt{SMASH} publically available code \cite{smash_code} necessary to reproduce our results related to deuterons.

 To reduce the effects of finite range interaction due to the geometric collision criterion, the testparticle method is used in \texttt{SMASH}. Specifically, at particlization the amount of particles is oversampled by factor $N_{test}$, and, at the same time, all cross sections in \texttt{SMASH} are reduced by factor $N_{test}$. We have shown previously (see Appendix of \cite{Oliinychenko:2018ugs}) for $N_{test} = 10$ that this helps to maintain the detailed balance for deuteron reactions with better than 1\% precision in an equilibrated box simulation. The effect of $N_{test}$ on our results is discussed further and is shown in Fig. \ref{fig:d_vs_t}.

\section{Deuteron production} \label{sec:results_deuteron}

\begin{figure}
    \centering
    \includegraphics[width=0.45\textwidth]{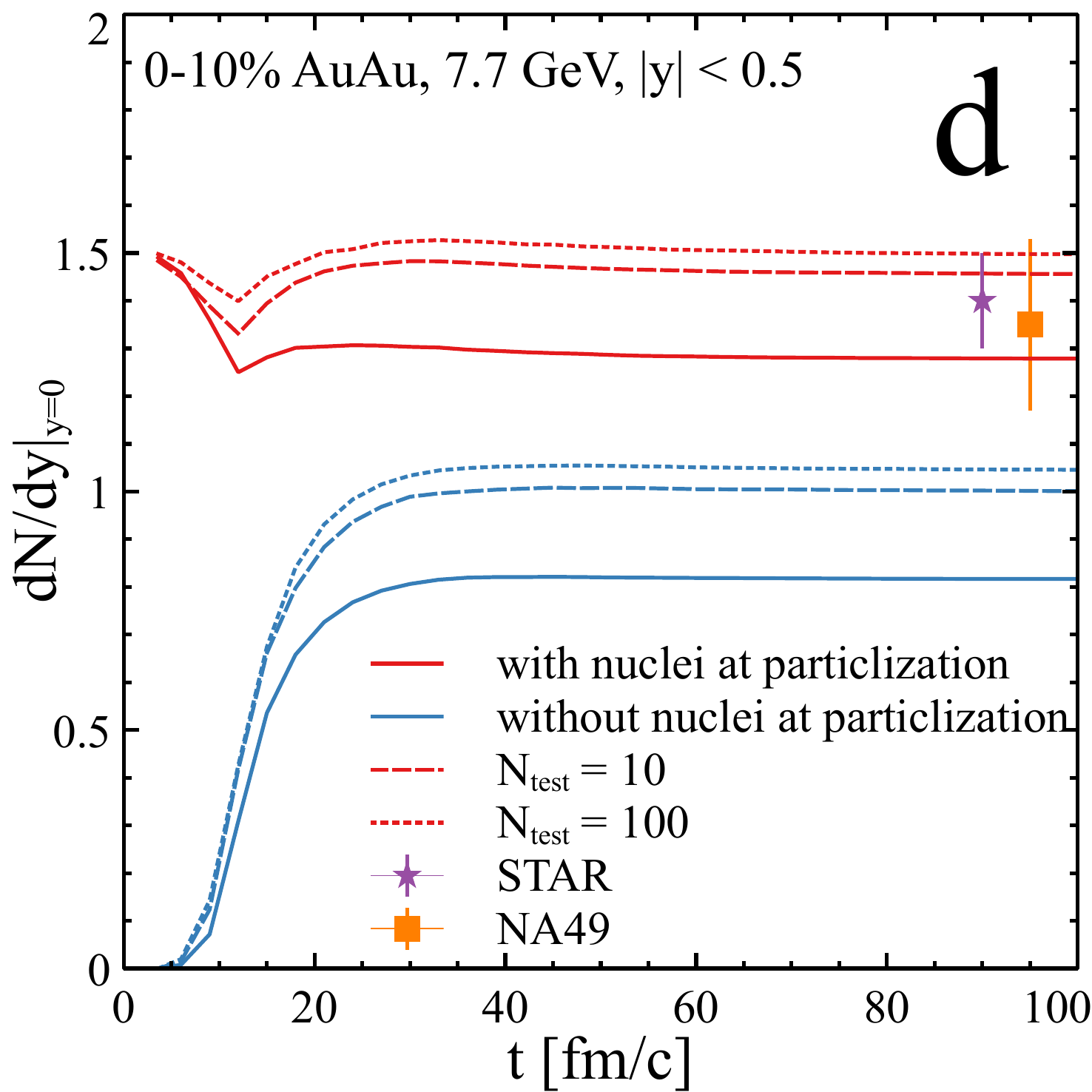}
    \caption{The deuteron yield as a function of time for 0-10\% central AuAu collisions at 7.7 GeV. Two cases are compared:  deuterons are sampled at particlization (red lines) and deuterons are not sampled (blue lines). Also the effect of testparticle number is shown: $N_{test} = 1$ (solid lines), $N_{test} = 10$ (dashed lines), and $N_{test} = 100$ (dotted lines). The experimentally measured yields by STAR \cite{Adamczyk:2017iwn} and NA49 \cite{Anticic:2010mp} are shown for comparison.}
    \label{fig:d_vs_t}
\end{figure}

Before comparing the deuteron production to experimental data, let us first explore its general features in our simulation. For this let us consider deuterons at our lowest energy (7.7 GeV) at midrapidity, $|y| < 0.5$. Similarly to \cite{Oliinychenko:2018ugs}, we try two scenarios: (i) deuterons are sampled at the particlization (ii) deuterons are not sampled at the particlization. In the second scenario  all deuterons are created in the hadronic afterburner. The deuteron yield in this case is around 30\% smaller than in case (i), and the data favors sampling of deuterons at particlization.

In Fig.~\ref{fig:d_vs_t} we also show the effect of the number of testparticles,  $N_{test}$  used in the simulation. Larger $N_{test}$ reduces the non-locality of our geometric collision criterion, to which deuterons seem to be rather sensitive because of the large production and destruction cross sections. The deuteron yield increases by almost 30\%, when $N_{test}$ is increased from 1 to 10. Further increase of $N_{test}$ does not change the deuteron yield significantly, as shown in Fig.~\ref{fig:d_vs_t}, therefore in our simulations we use $N_{test} = 10$.

In Fig. \ref{fig:d_vs_t} one can see that the deuteron yield does not change significantly over time in case when deuterons are sampled at particlization. At the particlization hypersurface (which in our case can also be called hadronic freeze-out surface, because stable hadron yields including resonance decay contributions are changing at most by 10\% in the hadronic afterburner) the deuteron yield is already close to the measured yield, however its transverse momentum spectra at this point correspond to a hadronic chemical freeze-out temperature. Later the deuteron momentum spectrum changes, but the yield stays approximately constant. We understand it as a result of deuteron being in relative equilibrium with nucleons: the amount of deuterons is determined by the amount of nucleons, which stays approximately constant. The relative equilibrium is kept mostly by $\pi d \leftrightarrow \pi p n$ reactions. These are the same features of deuteron production
that we observed in \cite{Oliinychenko:2018ugs} for PbPb collisions at a much higher 2.76 TeV energy. In Fig. \ref{fig:d_vs_t} there is a small initial dip in the deuteron yield as a function of time. We observed a similar dip in \cite{Oliinychenko:2018ugs} at 2.76 TeV. We attribute it to the fact, that we do not sample $d'$: although deuterons start in relative equilibrium with protons, it takes time to equilibrate $d$ and $d'$ together. The dip is, therefore, an unwanted, but luckily small, artifact of the fake $d'$ resonance.  

\begin{figure*}
    \centering
    \includegraphics[width=0.95\textwidth]{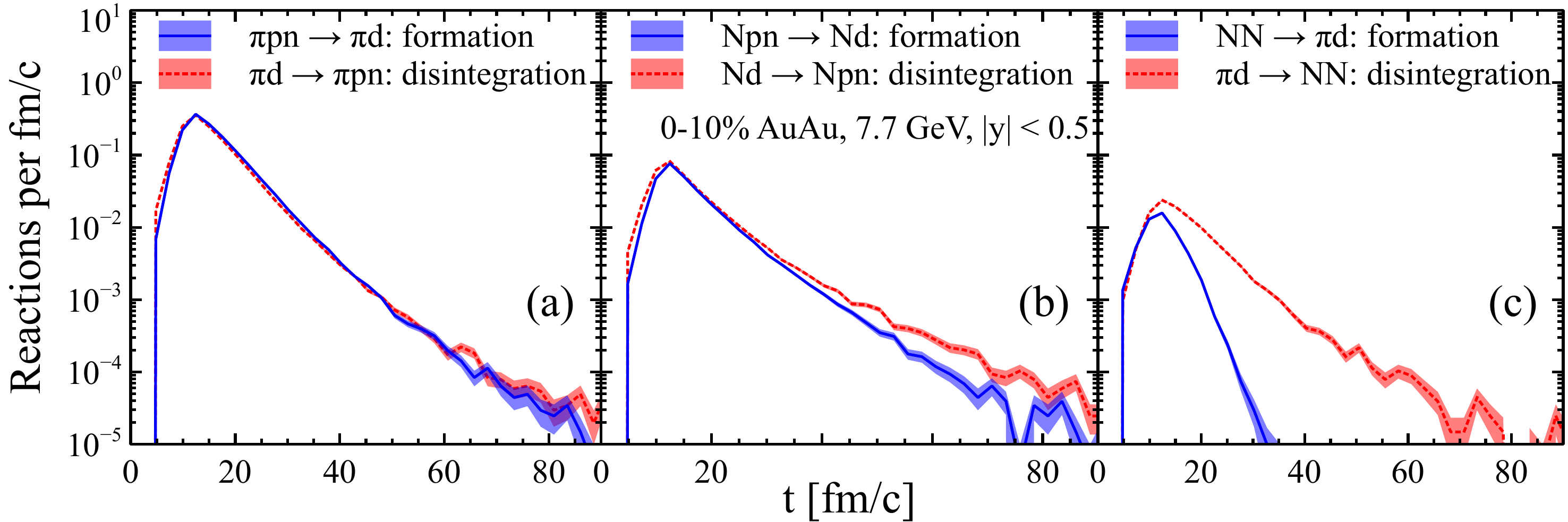}
    \caption{Event-averaged rates of reactions forming and disintegrating deuterons in central AuAu collisions at 7.7 GeV. For $\pi d\leftrightarrow \pi pn$ (left panel) and $N d\leftrightarrow N pn$ (middle panel) the forward and backward rates differ by 5-10\% at most between 10 to 40 fm/c. These reactions are close to being equilibrated. In contrast, $\pi d\leftrightarrow pn$ (right panel) has a much lower rate, it is not equilibrated, and destroys more deuterons than produces. However, the latter reaction contributes negligibly to the deuteron yield because of its low rate.}
    \label{fig:reactions}
\end{figure*}

 To further illustrate the picture described above, in Fig. \ref{fig:reactions} we show the reaction rates at midrapidity (rapidity of the reaction was computed from the total momentum of the incoming particles) at 7.7 GeV. One can see that the rates of forward and reverse $\pi d \leftrightarrow \pi p n$ almost coincide, the differences not exceeding 5\%. The $N d \leftrightarrow N p n$ reactions occur at several times lower rate than $\pi d \leftrightarrow \pi p n$, as one can see in Fig. \ref{fig:reactions}. It means that the $\pi d \leftrightarrow \pi p n$ reaction is dominant even at the energy as low as 7.7 GeV. In fact, we have observed in the separate pure transport simulation that at 7.7 GeV $\pi d \leftrightarrow \pi p n$ reactions alone are fast enough to drive deuteron into relative equilibrium with nucleons. Only below 4-5 GeV the $N d \leftrightarrow N p n$ reactions become more important than $\pi d \leftrightarrow \pi p n$, because for lower energies nucleon abundance increases, while pion abundance decreases. The $\pi d \leftrightarrow NN$ reactions are out of equilibrium, with deuteron destruction dominating at late time. However, their rate is negligible compared to $\pi d \leftrightarrow \pi np$ and the integrated rate over time is too small to influence the deuteron yield. The concept of deuterons in relative equilibrium is easy to misinterpret as deuterons being repeatedly formed and destroyed during the simulation. Such an interpretation is not correct, because at the energies considered here deuterons are rare particles. At midrapidity there are on average less than 2 reactions forming or destroying deuteron per event. For example, in a typical AuAu collision at 7.7 GeV one deuteron will be destroyed and one deuteron will be formed at midrapidity. The relative equilibrium one observes in Figs. \ref{fig:d_vs_t} and \ref{fig:reactions} emerges statistically after averaging over events.

Altogether, above we have established that our simulation behaves in a similar way from 7 up to 200 GeV, as at 2.76 TeV. Apriori it is not obvious that this behaviour should still be consistent with the experimental observables. It is not excluded that some new physical phenomena become important at 7--200 GeV, that did not play role at 2.76 TeV; it could be for example contributions from excited states of $^4\mathrm{He}$ \cite{Shuryak:2019ikv} (expected to be small for deuteron, here we just use them as an example) or a vicinity of the critical point.

\begin{figure*}
    \centering
    \includegraphics[width=0.32\textwidth]{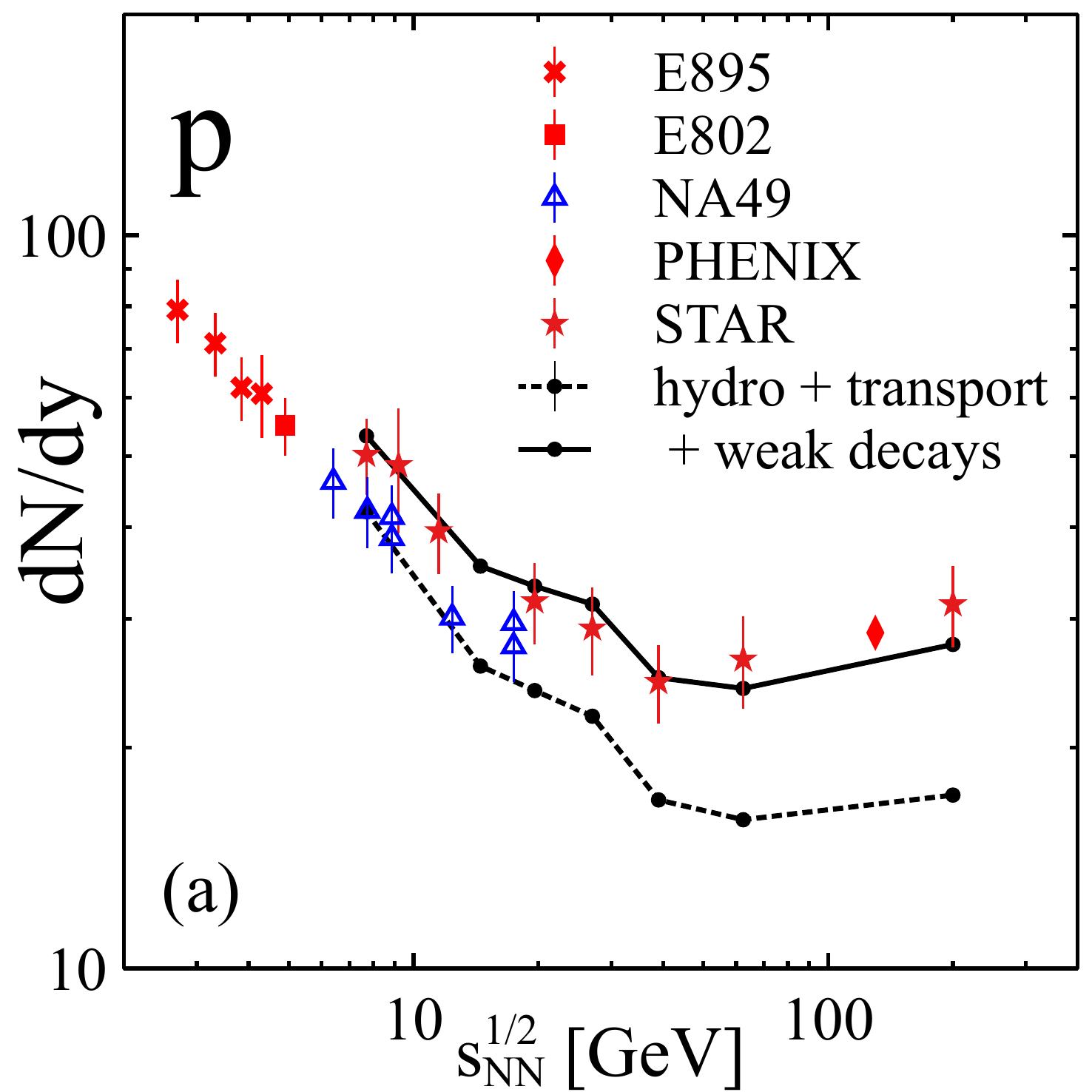}
    \includegraphics[width=0.32\textwidth]{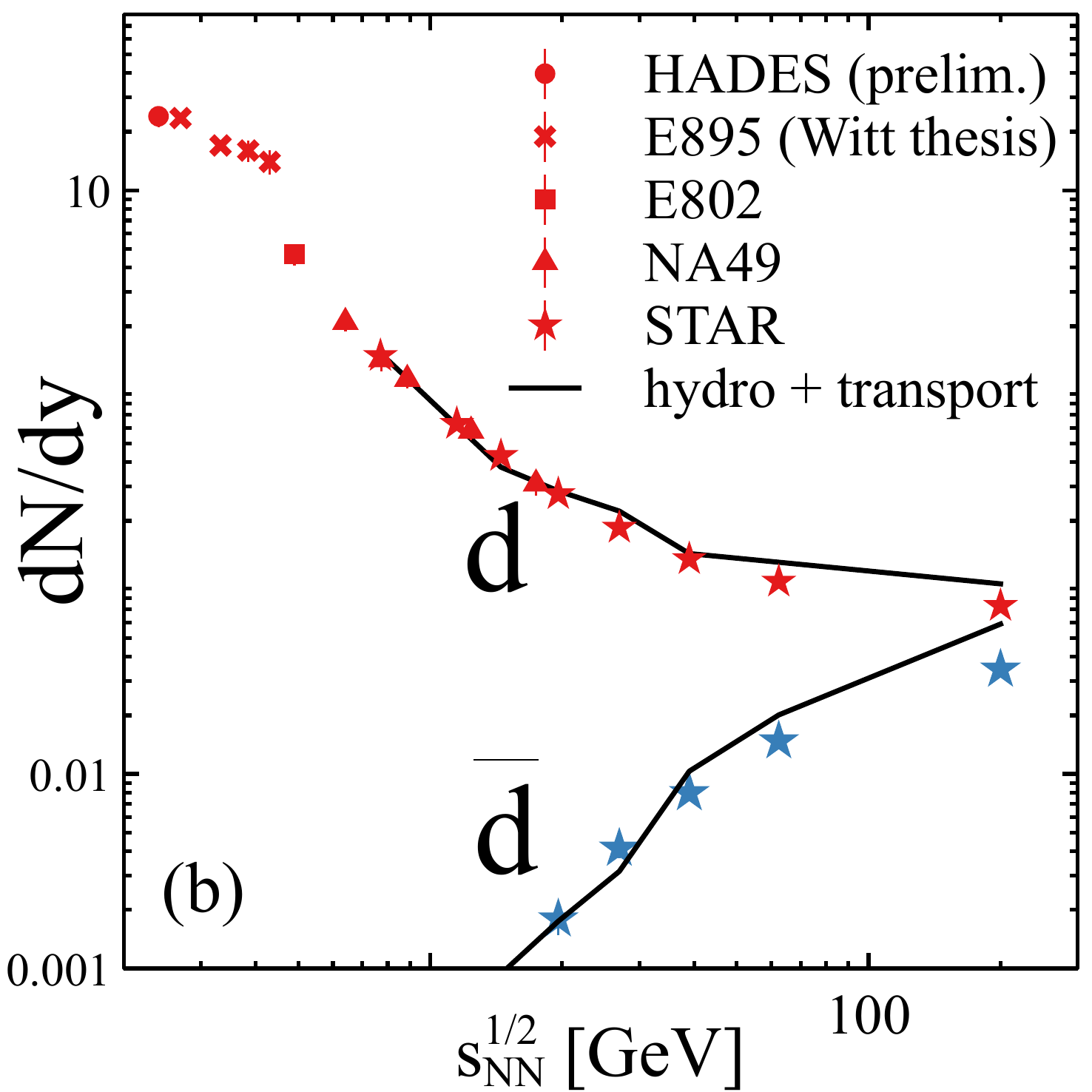}
    \includegraphics[width=0.32\textwidth]{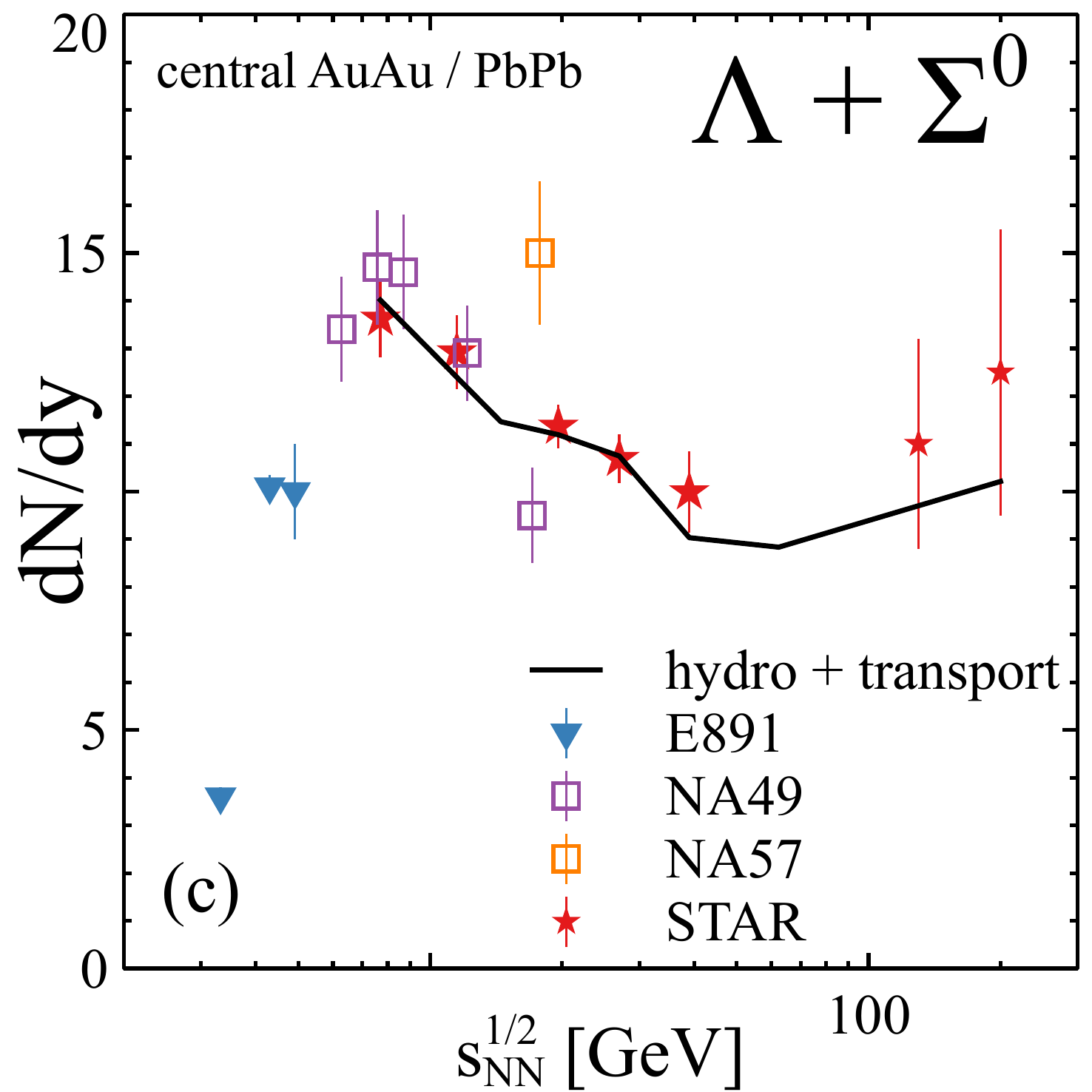}
    \caption{Proton, (anti-)deuteron, and $\Lambda$ yields at midrapidity in 0-10\% central AuAu collisions from \texttt{MUSIC} + \texttt{SMASH} simulation compared to data (proton: E895 \cite{Klay:2001tf}, E802 \cite{Ahle:1999in,Chen:1995af}, NA49 \cite{Alt:2006dk,Anticic:2010mp}, PHENIX \cite{Adcox:2003nr}, STAR \cite{Abelev:2009bw,Adamczyk:2017iwn}; $\Lambda$: E891 \cite{Ahmad:1991nv}, NA49 \cite{Alt:2008qm}, NA57 \cite{Antinori:2006ij}, STAR \cite{Adam:2019koz,Adler:2002uv,Adams:2006ke}; deuteron: HADES \cite{HADES_MLorentz_talk}, E895 \cite{Witt_thesis}, E802 \cite{Ahle:1999in}, NA49 \cite{Anticic:2016ckv}, STAR \cite{Adam:2019wnb}). In panel (a) proton yield data corrected for weak decays are shown with open blue symbols, while weak-decay-inclusive data are marked with full red symbols.}
    \label{fig:dn_dy}
\end{figure*}

\begin{figure*}
    \centering
    \includegraphics[width=0.32\textwidth]{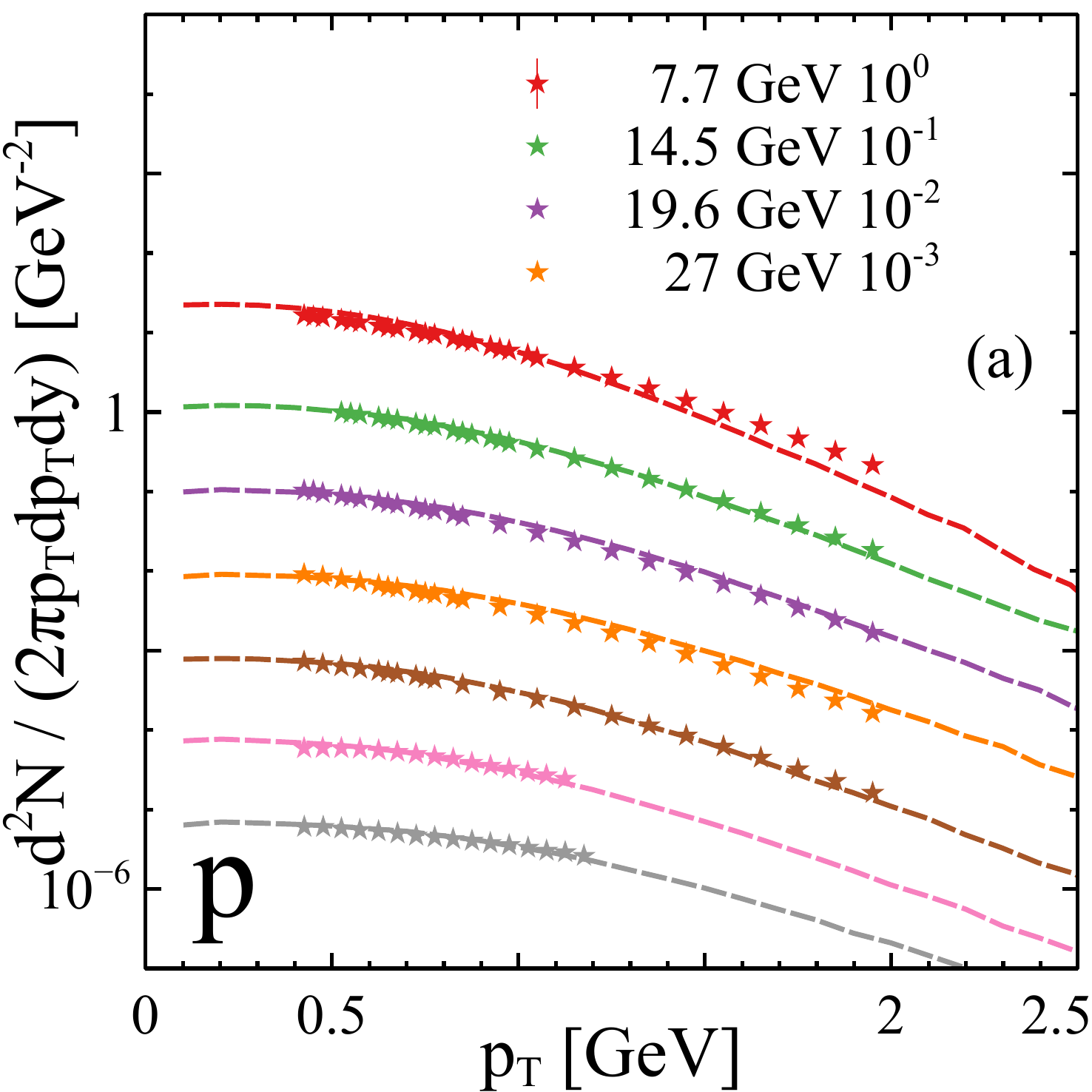}
    \includegraphics[width=0.32\textwidth]{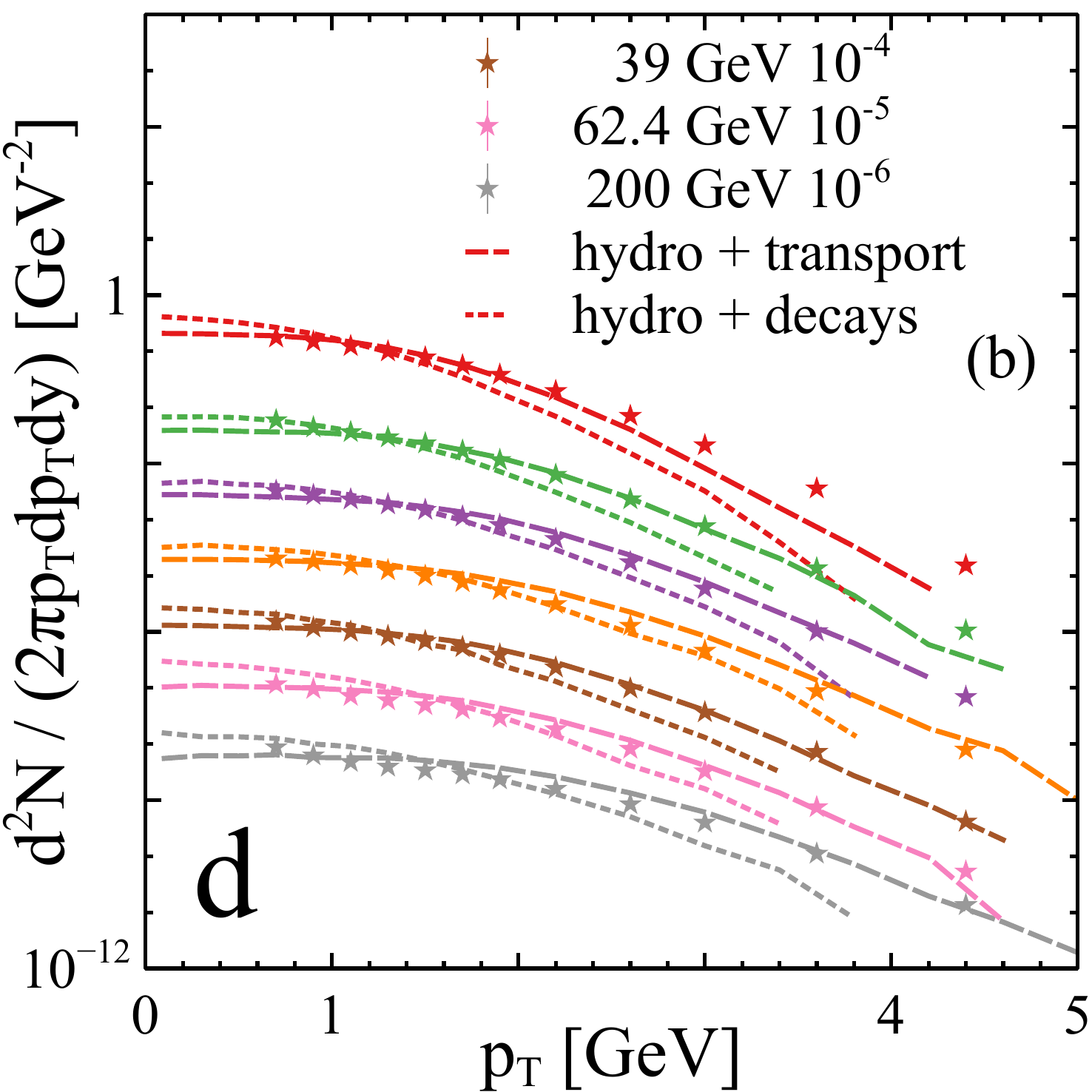}
    \includegraphics[width=0.32\textwidth]{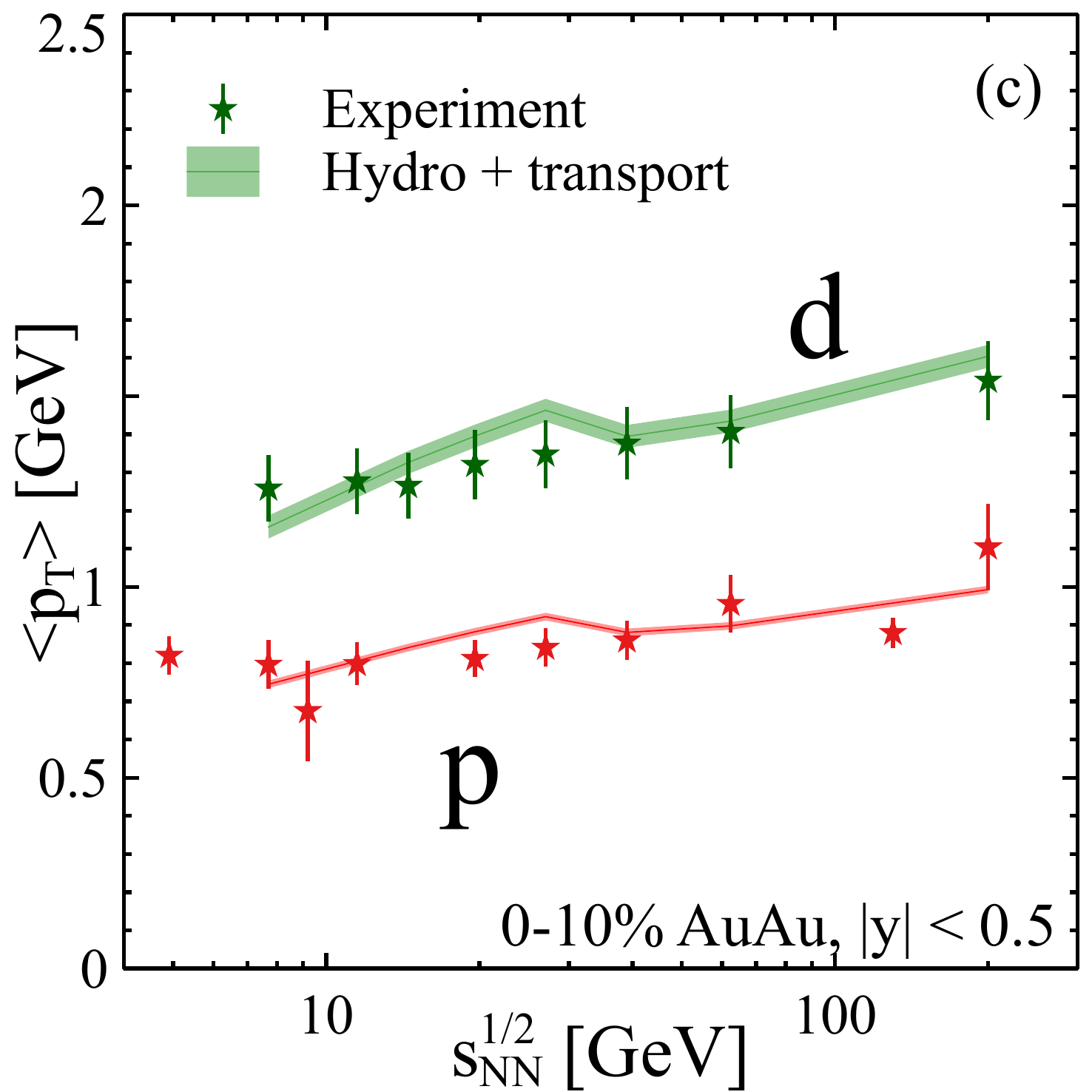}
    \caption{
    Proton (panel a) and deuteron (panel b) transverse momentum spectra and $\langle p_T \rangle$ (panel c) in 0-10\% central AuAu collisions at $\sqrt{s_{NN}} = 7.7-200$ GeV from our simulation are compared to data~\cite{Adamczyk:2017iwn,Adam:2019wnb}. Dashed lines correspond to the hydrodynamics + transport simulation, stars are experimental data points. Dotted lines in panel (b) are deuteron spectra at particlization. The difference between dashed and dotted lines demonstrates the effect of the afterburner for deuterons.}
    \label{fig:pT_spectra}
\end{figure*}

Already in \cite{Oliinychenko:2018ugs} we have noticed that a reasonable proton description is crucial for meaningful deuteron studies. Therefore, our first step in comparison of our simulation results to experiment is to test, if the hybrid \texttt{MUSIC} + \texttt{SMASH} approach is able to reproduce proton yields and transverse momentum spectra.  One caveat in such test is that proton yields measured by NA49 collaboration are corrected for weak decays \cite{Alt:2006dk,Anticic:2010mp}, while those measured by many other collaborations are not. Specifically, in STAR \cite{Adams:2003xp,Adamczyk:2017iwn,Abelev:2009bw}, PHENIX \cite{Adcox:2003nr} (although a correction for $\Lambda$ decays is available \cite{Adcox:2002au}), E895 \cite{Klay:2001tf}, E802 \cite{Ahle:1999in,Chen:1995af}, and preliminary HADES data proton yields and spectra \cite{HADES_MLorentz_talk} are not corrected for the weak decay feeddown. This causes an apparent disagreement between NA49 and STAR data shown in Fig. \ref{fig:dn_dy}. However, from the left panel of Fig. \ref{fig:dn_dy} one can see, that when the weak decays are included, the experimental results both from NA49 and STAR are described well by our approach. The proton yields with weak decays in Fig. \ref{fig:dn_dy} include all possible weak decays into protons, which comprise contributions from $\Lambda$, $\Sigma^0$, $\Sigma^+$, $\Xi^-$, and $\Omega$. This allows a fair comparison to the STAR data, because protons from STAR are truly inclusive with respect to weak decays \cite{Adamczyk:2017iwn}. To make sure that we correctly account for weak decays, we check the midrapidity yield of the $\Lambda$-hyperon. In Fig. \ref{fig:dn_dy} we show $\Lambda$ + $\Sigma^{0}$ yields for a fair comparison, because $\Sigma^{0}$  has a very short lifetime of $7.4 \cdot 10^{-20}$ s and decays with almost 100\% branching ratio as $\Sigma^0 \to \Lambda \gamma$ \cite{Patrignani:2016xqp}. This makes $\Sigma^0$ experimentally indistinguishable from $\Lambda$ in heavy ion collisions. The $\Lambda$ yields in Fig. \ref{fig:dn_dy}
do not include weak decay contributions from $\Xi$ and $\Omega$ baryons, both in our model and in experiment~\cite{Adam:2019koz}. We also reproduce the proton $p_{T}$ spectra rather well, as one can see in Fig. \ref{fig:pT_spectra}. The $p_T$ spectra are characterized comprehensively by the integrated yield $dN/dy$ and mean transverse momentum $\langle p_T \rangle$. In Figs. \ref{fig:dn_dy} and \ref{fig:pT_spectra} one can see that they are reproduced in our calculation for protons. A small cusp in proton and deuteron $\langle p_T \rangle$ at 19.6 GeV  originates from the fact that the starting time of hydrodynamics $\tau_0$ is tuned individually at each collision energy. While the $\langle p_T \rangle$ of $\Lambda$ is not shown, we have checked that the $\langle m_T \rangle$ is the same as that for protons within error bars, both in our model and in experiment. To sum up, as demonstrated  in Figs.~\ref{fig:dn_dy} and \ref{fig:pT_spectra} proton and $\Lambda$ yields, spectra, and $\langle p_T \rangle$ are described very well by our approach.

Furthermore, one can see in Fig. \ref{fig:dn_dy}, that the deuteron yields from different experiments \cite{Adam:2019wnb,Anticic:2016ckv,Ahle:1999in,HADES_MLorentz_talk,Witt_thesis}, as well as spectra and $\langle p_T \rangle$ are in good agreement with the \texttt{MUSIC} + \texttt{SMASH} simulations. We notice that wherever the proton spectrum in our model deviates from experiment, the deuteron spectrum qualitatively deviates in the same way. For example, at 7.7 GeV, where our description of proton spectra is the least accurate (to improve it the initial longitudinal baryon density profile in the hydrodynamics has to be considered more carefully) and over(under)shoots the data, the deuteron spectrum also over(under)shoots. Therefore we conjecture that if we tune the model to reproduce proton observables even better, the deuteron description will also improve.

The reactions involving anti-deuterons in \texttt{SMASH} are the CPT-conjugated deuteron reactions with the same cross sections. Consequently, proton and deuteron yields and spectra are connected in the same way as anti-proton and anti-deuteron yields and spectra. Just like for deuterons, pion catalysis $\pi \bar{d} \leftrightarrow \pi \bar{n} \bar{p}$ is the most important reaction for anti-deuteron production, and it leads to a reasonable description of the anti-deuteron yields, as shown in Fig. \ref{fig:dn_dy}. At 62.4 and 200 GeV one can see in Fig.~\ref{fig:dn_dy} that the anti-deuteron yield overshoots in our model. This is mainly because the anti-proton yield overshoots. A better description of protons and anti-protons at 62.4 and 200 GeV will require simultaneous fine-tuning of the initial hydrodynamical baryon density profile (aka baryon stopping) together with the switching energy density $\epsilon_{sw}$.

As we have shown, the quality of the model description of proton and  deuteron spectra are strongly related. This suggests that ratios of these spectra should be described even better than the yields. Therefore we construct the so called $B_2$ ratio of the spectra, which plays an important role in coalescence models:
\begin{eqnarray} \label{eq:B2definition}
B_2(p_T) = \left(\frac{dN_d(p^d_T / 2)}{2 \pi p_T dp_T dy}  \right) / \left( \frac{dN_p(p^p_T)}{2 \pi p_T dp_T dy}  \right)^2 \,.
\end{eqnarray}

In coalescence models this ratio is roughly inversely proportional to an emission volume if this emission volume is much larger than the size of the deuteron, which is the case here. The $B_2$ ratio is known both theoretically and experimentally to grow with $p_T$, consistent with the volume from femtoscopic measurements \cite{Adamczyk:2014mxp} decreasing with $p_T$. The $B_2$ ratio is also known to decrease with increasing collision energy, again consistent with the increase of the volume from femtoscopic measurements with the energy. However, two non-trivial features are present in the STAR measurement of $B_2$ \cite{Adam:2019wnb}: (i) the experimentally measured $B_2(\sqrt{s}, p_T/A = 0.65 \gev)$ has a broad minimum at 20--60 GeV; (ii) the anti-deuteron $B_2^{\bar{d}}(\sqrt{s}, p_T/A = 0.65 \gev)$ is smaller than the deuteron $B_2$. In our previous work~\cite{Oliinychenko:2018ugs} we speculated  that the minimum of $B_2$ might be connected to a switch of the dominant deuteron production mechanism from $\pi p n \leftrightarrow \pi d$ to $N n p \leftrightarrow N d$ reactions. However, as shown in this work this conjecture is not supported by our calculations, because the $\pi p n \leftrightarrow \pi d$ reaction is dominating all the way down to $\sqrt{s}=7.7\gev$, which is well below the location of the minimum in $B_2$. Therefore,  let us inspect the behaviour of $B_2$ closer and suggest a possible  explanation, why it exhibits the aforementioned minimum.

\begin{figure*}
    \centering
    \includegraphics[width=0.32\textwidth]{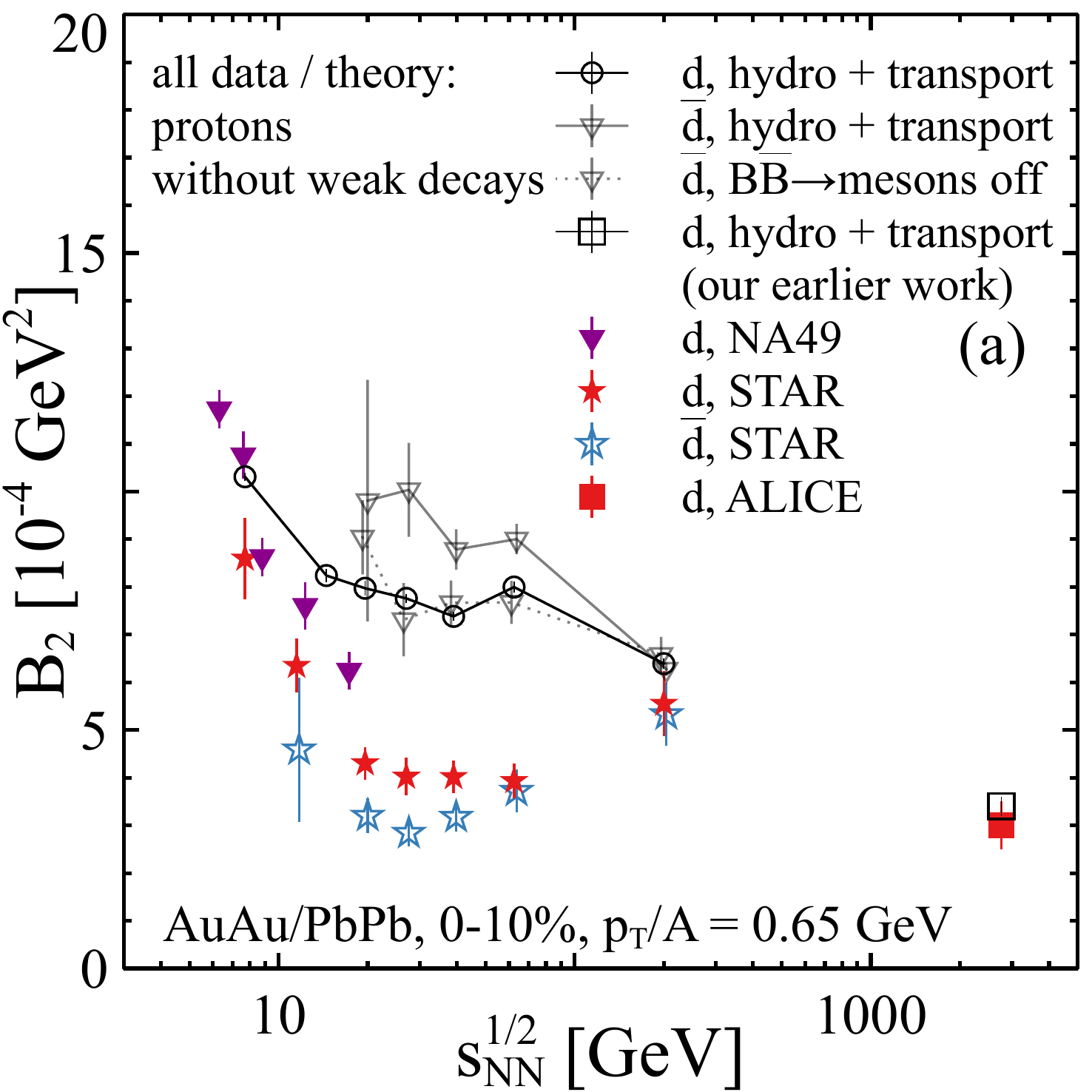}
    \includegraphics[width=0.32\textwidth]{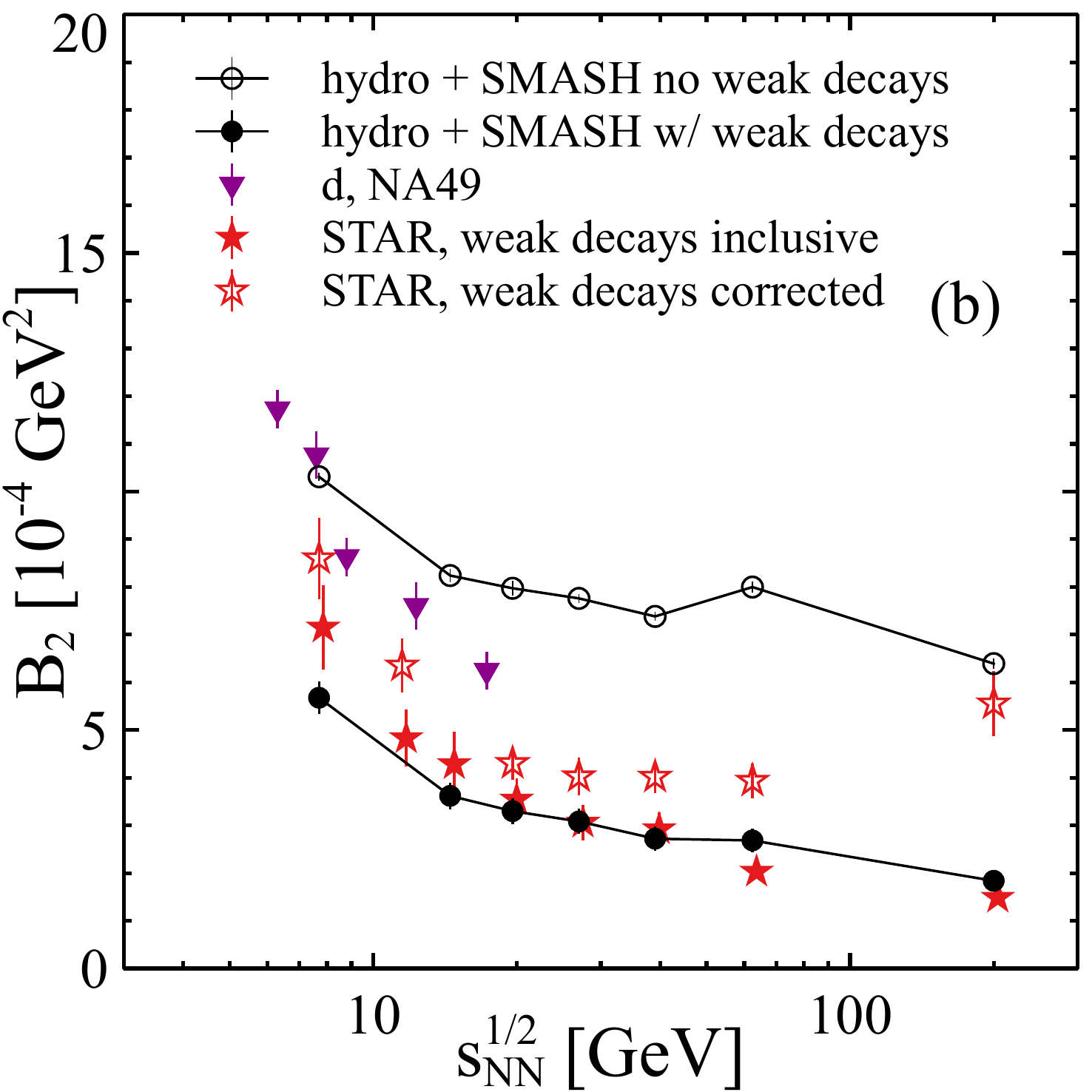}
    \includegraphics[width=0.32\textwidth]{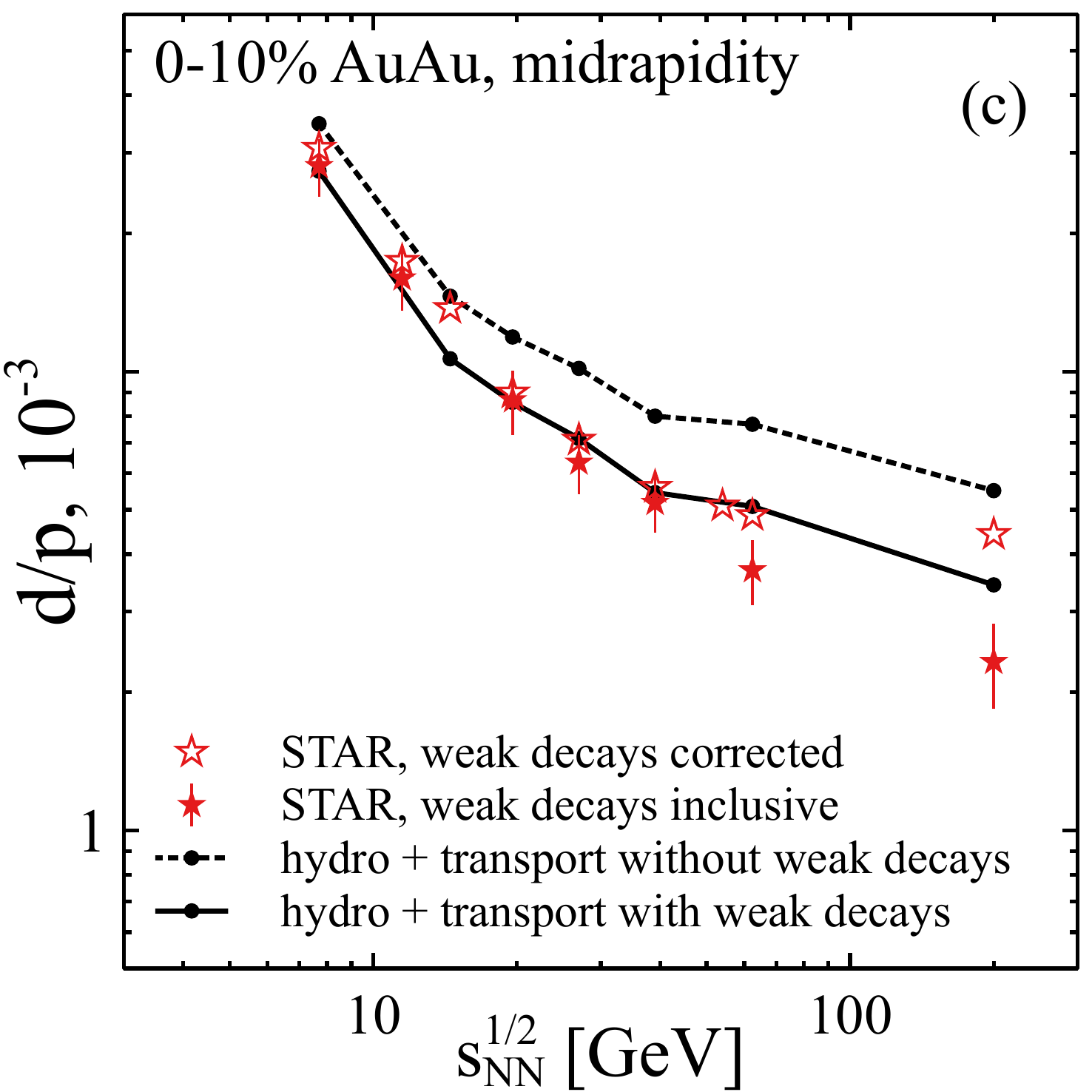}
    \caption{Comparison of the measured weak-decay-corrected $B_2(\sqrt{s})$ at $p_T/A$ = 0.65 GeV (panel a), weak-decay-inclusive $B_2$ (panel b), and deuteron to proton ratio (panel c) to our simulations. We tend to describe weak-decay-inclusive observables well, in contrast to poorly described weak-decay-corrected observables. }
    \label{fig:B2}
\end{figure*}

We find that the shape of the transverse momentum dependence of $B_2(p_T)$ is similar for all considered energies and matches the experiment rather well.
However, comparing the magnitude of $B_2(\sqrt{s}, p_T/A = 0.65 \gev)$ our simulation significantly overestimates the experimental values in the  energy range of $20 \gev < \sqrt{s} < 60 \gev$, where the minimum is located (see Fig. \ref{fig:B2}). This discrepancy is surprising given that we reproduce proton, $\Lambda$, and deuteron spectra rather well; after all $B_2$ is nothing but the ratio of the spectra. Investigating this closer we find that it is the weak decays that play a crucial role in this discrepancy. Indeed, if we compute $B_2$ by dividing the deuteron spectrum from STAR over the weak-decay-inclusive proton spectrum from STAR, we reproduce this weak-decay inclusive $B_2$ rather well, see Fig. \ref{fig:B2}b. This shows that the contribution from weak decays is much larger in our model than the weak decay correction in the STAR data. This is exacerbated by the fact that in the definition of $B_2$ the proton spectrum, and thus the weak decay correction, enters in square. Furthermore, comparing the $B_2$ ratio from STAR with and without weak decays we find that the $B_2$ ratio with weak-decay-inclusive protons does not exhibit a minimum, whereas that with weak-decay-corrected protons does show minimum (see Fig.~\ref{fig:B2}b). This suggests that the minimum structure in the energy dependence of $B_2^{\bar{d}}(\sqrt{s}, p_T/A = 0.65 \gev)$ may originate from the  weak decay corrections. To further explore the effect of weak decays we consider the the $d/p$ ratio (see Fig.~\ref{fig:B2}c). Again, our model calculation reproduces the weak-decay-inclusive $d/p$ ratio rather well, but not the preliminary weak-decay corrected $d/p$ ratio from STAR \cite{Zhang:2020ewj}.

\begin{figure}
    \centering
    \includegraphics[width=0.45\textwidth]{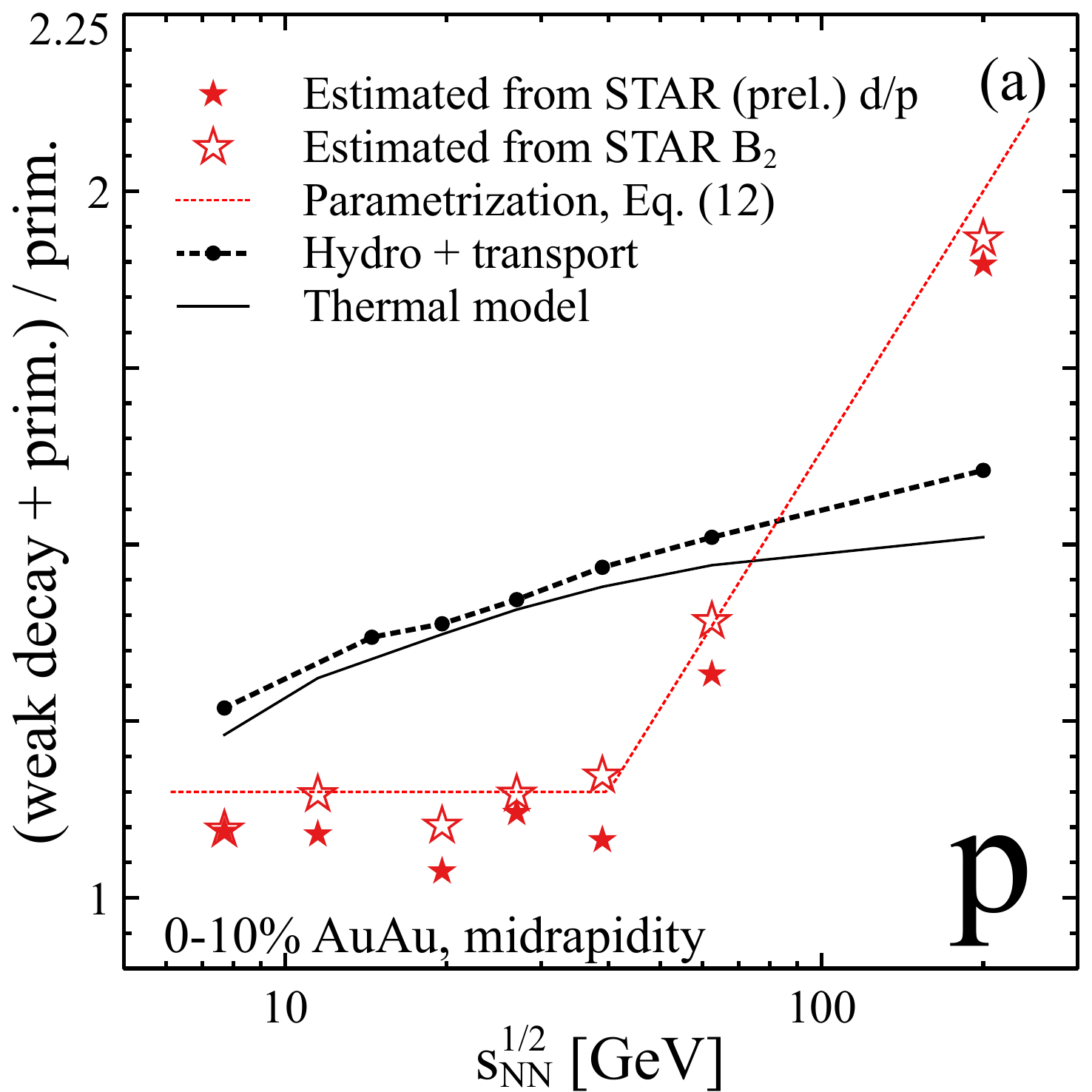}\\
    \includegraphics[width=0.45\textwidth]{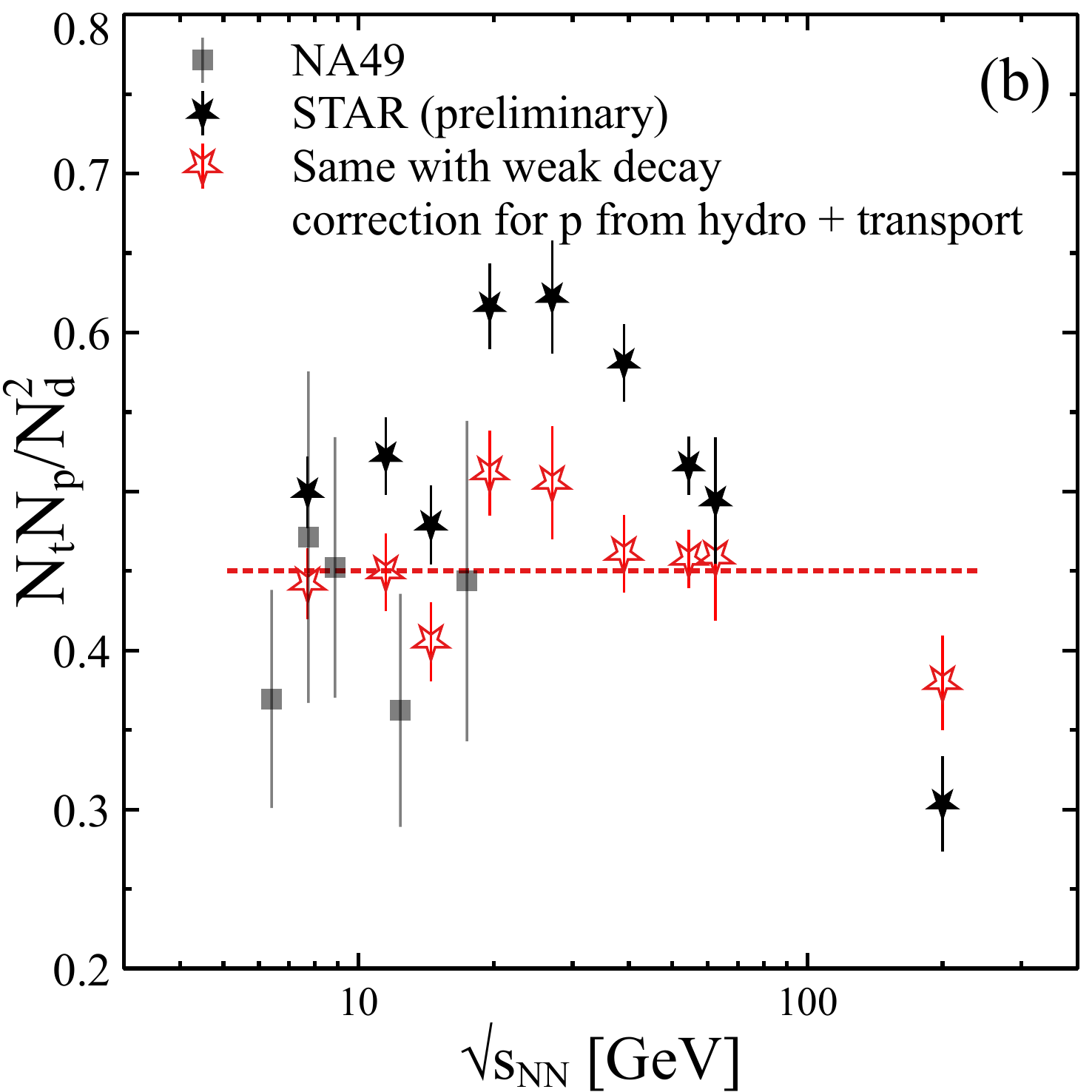}
    \caption{Revision of weak decay corrections. Panel (a): STAR weak decay corrections (red stars) are extracted in two ways from the available publications, from $d/p$ ratio \cite{Zhang:2020ewj} and from $B_2$ \cite{Adam:2019wnb}. It is compared to the weak decay corrections in our model (dashed line) and the thermal model (solid line). Dotted line is our approximate parametrization of STAR weak decay correction according to Eq. (\ref{eq:weak_decay_parametrization_STAR}). Panel (b): demonstration of how the measured $\frac{N_t N_p}{N_d^2}$ yield ratio would change if weak decay corrections resulting from our model were adopted.}
    \label{fig:weak_decay}
\end{figure}

Since our model describes both proton and $\Lambda$ yields well, one is led to the conjecture that the weak decay corrections to the measured proton yields might be underestimated. Our conjecture is inspired by our model, but it also has some model-independent support from experimental data. First, we notice in Fig.~\ref{fig:B2} that at the energies where STAR and NA49 data intersect, the weak-decay-corrected $B_2$ from NA49 is always higher, even though the Pb+Pb collision system is slightly larger than the Au+Au system of the STAR measurement. Since $B_2$ scales with the inverse size of the system, one would expect the $B_{2}$ ratio obtained by NA49 to be below that measured  by STAR. If, on the other hand, the weak decay correction to proton yields were larger in the STAR data, it would improve the agreement between STAR and NA49 results for $B_2$. Second, one can estimate the weak decay correction in a data-driven way from the recent STAR measurements of strange particle production. Let us consider such an estimate at 39 GeV. The measured yield of $\Lambda + \Sigma^0$ is around $dN^{\Lambda}/dy \approx 10$, see Fig. \ref{fig:dn_dy}c. Let us assume that the $\Sigma^+$ yield, which is not measured, is approximately equal to the $\Sigma^0$ and $\Lambda$ yields. This assumption is mainly motivated by the thermal model, where the yields are determined by hadron masses, which are close for $\Sigma^+$, $\Sigma^0$, and $\Lambda$. Additional contribution from $\Xi$ and $\Omega$ decays to $\Lambda$ constitutes around 10-20\% of the $\Lambda$ yield. Taking into account the branching ratios $BR(\Lambda \to p \pi^-) \approx 0.63$ and $BR(\Sigma^+ \to p \pi^0)\approx 0.52$, we obtain the yield of protons from weak decays $dN^{p-weak}/dy \approx 9-11$. Therefore, at 39 GeV at midrapidity around 20 protons per event are prompt and approximately  10 originate from weak decays. The weak decay correction coefficient is thus $\approx 30/20 = 1.5$. The STAR estimate is roughly 1.15--1.25,  both for the $B_2$ and $d/p$ ratio, see Fig. \ref{fig:weak_decay}. We note, that the weak decay correction estimate in \cite{Adam:2019wnb} is not data-driven, but involves the UrQMD model, and may  possibly be  model dependent. Needless to say, that  a data-driven weak decay correction would be beneficial to understand the $B_2(\sqrt{s})$ behaviour.

To sum up, our results as well as data-driven analysis suggest that the observed minimum in $B_2(\sqrt{s})$ may originate from the weak decay corrections of proton spectra. Moreover, there are both theoretical and experimental indications that these weak decay corrections are underestimated in \cite{Adam:2019wnb}. If these indications are true, then it has intriguing consequences, which we discuss next.

This work was largely inspired by the study \cite{Sun:2018jhg}, which relates the yields of light nuclei to spatial fluctuations of nucleon densities. Specifically, spatial nucleon density fluctuations are connected to the $\frac{N_t N_p}{N_d^2}$. This ratio has been measured by STAR recently \cite{Zhang:2020ewj} and it exhibits a peak, which might be a signal of the enhanced nucleon density fluctuations and therefore potentially a critical point. One can see the preliminary STAR data  in Fig. \ref{fig:weak_decay}. The proton yields in this measurement are corrected for weak decays. Suppose our conjecture about weak decay corrections turns out to be correct and the corrections have to be re-evaluated. What will the corrected $\frac{N_t N_p}{N_d^2}$ ratio be? To answer this question quantitatively we extract the ratio of total to non-weak-decay (prompt) protons from STAR data in two ways: from the published $B_2$ and from preliminary $d/p$ ratio, see Fig. \ref{fig:weak_decay}. These two ways do not have to give necessarily identical results. Indeed, in case of $d/p$ ratio the relevant observable is the $p_T$-integrated proton yield, while in $B_2$ it is $p_T$-differential yield at proton $p_T$ = 0.65 GeV. However, one can clearly see in Fig. \ref{fig:weak_decay} that the two ways are in agreement. For convenience we parameterize the STAR weak decay correction $x$ --- total proton yield over primordial proton yield without weak decays --- as
\begin{eqnarray} \label{eq:weak_decay_parametrization_STAR}
x_{STAR} =& \nonumber \\ = 1.15 
+& \frac{\log \sqrt{s} - \log 40}{\log 200 - \log 40} (2 - 1.15) \, \theta(\sqrt{s} - 40) \,,
\end{eqnarray}
where $\sqrt{s}$ is the collision energy in GeV. 
Our correction estimated from the hydro + transport approach (which reproduces the STAR experimental yields) can be approximately parametrized as
\begin{eqnarray} 
x_{model} = 1.3 + \frac{\log \sqrt{s} - \log 7.7}{\log 200 - \log 7.7} (1.6 - 1.3).
\end{eqnarray}
We have checked that the weak decay correction in our hydro + transport simulation closely resembles that of the thermal model using Thermal-FIST package \cite{Vovchenko:2019pjl}, see Fig. \ref{fig:weak_decay}. Similarly to our model, the thermal model describes both proton and $\Lambda$ yields rather well at STAR BES energies (7.7 -- 200 GeV). This supports our argument and allows to test it with a much simpler thermal model setup. Indeed, our calculation of weak decay correction for proton yields is in agreement with the \textit{thermal model} calculation presented by the STAR collaboration in Ref. \cite{Adamczyk:2017iwn}: we obtain $(N^{p}_{primordial} + N^{p}_{weak}) / N^{p}_{primordial} = 1.23$ at 7.7 GeV (Fig. \ref{fig:weak_decay}a) or $N^p_{weak} / (N^{p}_{primordial} + N^{p}_{weak})= 18.7$\%, consistent with 18\% in Ref. \cite{Adamczyk:2017iwn}; at 39 GeV we obtain $(N^{p}_{primordial} + N^{p}_{weak}) / N^{p}_{primordial} = 1.44$ or $N^p_{weak} / (N^{p}_{primordial} + N^{p}_{weak})= 30.6$ \%, again consistently with the 29\% STAR has obtained in their estimate,  Ref. \cite{Adamczyk:2017iwn}.

After correcting the preliminary STAR data for the $\frac{N_t N_p}{N_d^2}$ ratio by the factor $x_{model} / x_{STAR}$, we observe that the peak in the $\frac{N_t N_p}{N_d^2}(\sqrt{s})$ dependence becomes much less pronounced. In fact, the ratio scaled in this way becomes more consistent with a constant value predicted by the coalescence models in absence of non-trivial effects, such as enhanced nucleon density fluctuations. Finally we would like to underline that our analysis is by no means conclusive. We simply want to draw attention to the ``technical'' issue  of weak decay corrections  pointing  out that it may influence the interpretation of the data in a profound way.

We have already mentioned that the $B_2$ ratio for anti-deuterons measured by STAR is smaller than that for deuterons. In Fig. \ref{fig:B2}a one can see that our simulation produces the opposite trend: the $B_2$ of anti-deuterons is larger than the $B_2$ of deuterons. It appears that baryon annihilation, $B\bar{B} \to \mathrm{mesons}$  plays a prominent role in that difference. In \texttt{SMASH} this reaction is not balanced: the annihilation $B\bar{B} \to \mathrm{mesons}$ is allowed, but the reverse process is not possible. After switching off the $B\bar{B} \to \mathrm{mesons}$ annihilation reaction we obtain a lower $B_2(\bar{d})$ (see Fig. \ref{fig:B2}a), while $B_2(d)$ remains the same except for 62.4 and 200~GeV, where $B_2(d)$ is also slightly reduced. As one can see in Fig. \ref{fig:B2}a, without baryon annihilation $B_2(\bar{d}) \approx B_2(d)$ within statistical error bars. Thus it seems that the experimentally measured difference of $B_2$ for deuterons and anti-deuterons may be due to baryon-antibaryon annihilations. However, other possibilities are not excluded, for example, the effect of a weak decay correction, which is larger for anti-protons than for protons.
 
 \section{Summary and discussion} \label{sec:summary}

In summary,  the results of this work and \cite{Oliinychenko:2018ugs}, based on hydrodynamics plus transport simulations show that it is possible to reproduce deuteron yields and spectra at energies 7--2760 GeV using pion catalysis reactions $\pi d \leftrightarrow \pi p n$ with large cross sections obeying the detailed balance principle. One important detail is that the underlying hydrodynamical simulation has to be tuned to reproduce protons and Lambdas well, which we successfully accomplish. The conclusions from \cite{Oliinychenko:2018ugs} regarding deuteron staying in relative equilibrium with nucleons and its yield being almost constant starting from hadronic chemical freeze-out are still valid down to a collision energy of $\sqrt{s} = 7\gev$. This also explains the apparent puzzle why the deuteron yield is determined at chemical freeze-out while their spectra correspond to the kinetic freeze-out. At lower energies the deuteron production mechanism is expected to change: $N d\leftrightarrow N pn$ reactions will start to dominate.
 
Analysing the $B_2$ ratio, which is a ratio of  the deuteron spectrum over square of the proton spectrum (Eq. \ref{eq:B2definition}), we realized that weak decay corrections to the proton spectrum play a significant role. In particular we found that the observed minimum in $B_2(\sqrt{s})$ is most likely related to the weak decay corrections. Furthermore, we noticed that weak decay corrections to the proton yield also affect the interpretation of the $\frac{N_t N_p}{N_d^2}$ ratio, which exhibits an unexplained peak as a function of the collision energy. We also demonstrated that the difference between the measured $B_2(\bar{d})$ of anti-deuterons and $B_2(d)$ of deuterons might be related to $B\bar{B}$ annihilations. A recent work~\cite{Kittiratpattana:2020daw} explains the difference between $B_2$ of deuterons and antideuterons as a consequence of the fact that due to annihilations anti-deuterons are emitted from a smaller volume. Although we agree with \cite{Kittiratpattana:2020daw} that baryon-antibaryon annihilations are likely responsible for the $B_2$ difference, we notice that a smaller emission volume should result in a larger, not smaller $B_2$.
 
It seems that we have reached a satisfactory understanding of proton and deuteron production across the STAR beam energy scan energies. The next step is to consider the production of $A = 3$ nuclei: triton, helium-3, and possibly hypertriton.
Unfortunately, the extension of our method to $A= 3$ nuclei requires an additional fake resonance, $t'$. It can be avoided,  however, through implementing $2 \leftrightarrow 3$ reactions via stochastic rates method. This work is currently in progress.
 
 Another possible extension of the present work is to consider the role of the mean field nuclear potentials on the light nuclei production. Since the light nuclei are mostly formed at the late stage of collision, their yields may be sensitive to the nuclear mean fields at few normal nuclear densities and below. Therefore, it would be interesting to verify to which extent (if any) the nuclear matter liquid-gas phase transition influences the light nuclei yields.

\begin{acknowledgements}
We thank K. Declan and J. Klay for sharing E895 proton and deuteron yields and for their comments regarding the data. D. O. thanks J. Mohs and H. Elfner for fruitful discussions. D. O. and V. K. were supported by the U.S. Department of Energy, Office of Science, Office of Nuclear Physics, under contract number DE-AC02-05CH11231.
C.~S. was supported in part by the U.S. Department of Energy (DOE)  under grant number DE-SC0013460 and in part by the National Science Foundation (NSF) under grant number PHY-2012922.
This work also received support within the framework of the Beam Energy Scan Theory (BEST) Topical Collaboration.
Computational resources were provided by NERSC computing cluster, the high performance computing services at Wayne State University, and by Goethe-HLR cluster within the framework of the Landes-Offensive zur Entwicklung Wissenschaftlich-\"Okonomischer Exzellenz (LOEWE) program launched by the State of Hesse. D.O. was supported by the U.S. DOE under Grant No. DE-FG02-00ER4113.
\end{acknowledgements}

\bibliography{inspire,noninspire}

\end{document}